# Network Security Threats and Protection Models

**Amit Kumar** and **Santosh Malhotra**

*kamit5@iitk.ac.in, mahotsan@iitk.ac.in*

Department of Computer Science and Engineering

Indian Institute of Technology Kanpur

Kanpur, Uttar Pradesh 208016, India

## 1. INTRODUCTION

In a brave new age of global connectivity and e-commerce, interconnections via networks have heightened, creating for both individuals and organizations, a state of complete dependence upon vulnerable systems for storage and transfer of information. Never before, have so many people had power in their own hands. The power to deface websites, access personal mail accounts, and worse more the potential to bring down entire governments, and financial corporations through openly documented software codes. This paper discusses the possible exploits on typical network components, it will cite real life scenarios, and propose practical measures that can be taken as safeguard. Then, it describes some of the key efforts done by the research community to prevent such attacks, mainly by using Firewall and Intrusion Detection Systems.

## 2. NETWORK SECURITY THREAT MODELS

Network security refers to activities designed to protect a network. These activities ensure usability, reliability, and safety of a business network infrastructure and data. Effectual network security focuses on a variety of threats and hinders them from penetrating or spreading into the network. Figure 1 shows some of the typical cyber attack models.

The most common threats include:

- Trojan horses and spyware (spy programs)
- DOS (Denial of service attacks)
- Data interception and theft

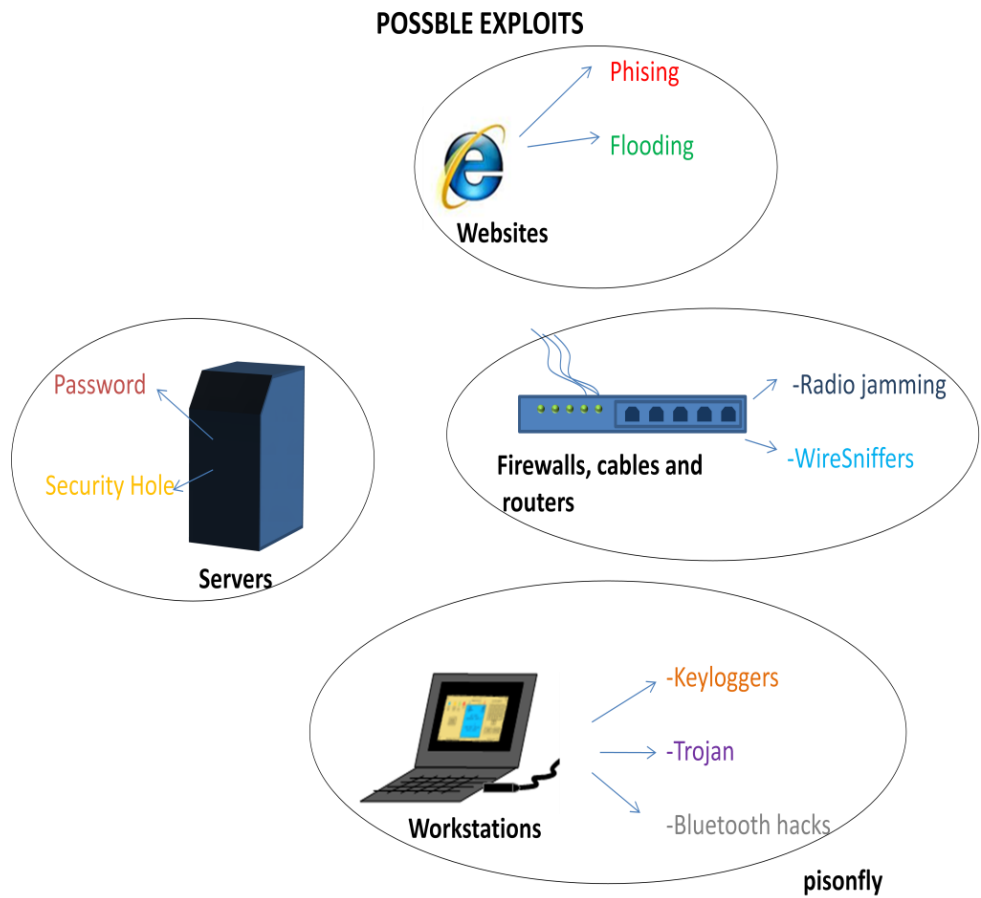

**Figure 1:** Common Cyber Attack Models

    a. **FLOODING**

In 1998, an American elite group, "The Digital Disturbance Theater"    with Floodnet, an application set to halt the Mexican president's webpage (for political reasons). Floodnet is a java applet that automates the "refresh" button to click repeatedly. Sufficient users online would run the application and hence cause the site's server to continuously refresh until saturation and thus halt and disable the webpage. An attacker has used similar applications to take into hostage commercial websites in exchange for ransom. It is advisable for an organization to have for emergency, a savvy security expert (White-hat hacker); seeing that web technology is dynamic, with the ever changing trends in web scripting languages and browser configurations.

b. **KEYLOGERS**

These are simple software codes that exploit what we call 'hooks' on a computer's kernel. Hooks capture vital hardware traffic like Keystrokes and mouse movements. Software based Key loggers are programmed to capture any button stroke you type on the keyboard and save words as a text file. That includes all private information you type like Passwords, Google searches, Credit card number, emails, to name but a few. Regularly updating of the Antivirus is a sure way to beat this. Let it also be known that Hardware key loggers exist, masquerading as flash disks. USB password applications should deter such.

c. **TROJANS**

An experienced programmer is capable of creating a Trojan, a concealed application that runs in the background. A Trojan allows a hacker to become a ghost user on your PC/Workstation. They monitor when your computer is online to deliver captured keystroke log files to their preferred address. Hackers can always come back and upload a malicious code via the Trojan. Such a code maybe the one that kills your antivirus program after which, it takes your snap via webcam or taps into your office conversations from your laptop microphone. Trojans come tucked away neatly on pirated software and the so-called cracks we all like to use. As the adage goes, it is difficult to cheat an honest person. The converse is true for those who would escape this pitfall. Let them invest in genuine software.

d. **BLUETOOTH**

Bluetooth is emerging as a versatile networking technology connecting workstations to printers, smart phones etc. I see potential for mischief; where data could be wirelessly intercepted for malicious use. Such technology is currently non-existent, to the best of my knowledge, but nonetheless, a practical possibility.

e. **PHISING**

This is when emails appearing to come from well known organizations pop up on your browser, sending you links and requesting for private information like credit card numbers, account passwords or congratulating you for winning. Watch out for that nice email from a website you do not even have an account with.

Look-alike websites are also not uncommon. They will have you login and 'refill' your personal details; after which they can make online purchases under your name or if they be diabolical enough, they will lock you out of your own account. (I lost my yahoo account that way). Numerous cyber security forums and workshops exist where one can always learn ways to have an edge over scammers and keep your business team informed

f. **RADIO JAMMING**

This can be a rare DOS (Denial of Service) technique to disrupt information flow in a wireless router network, accomplished by use of noise-generating radio devices. However, special Equipment exist, that can be used to track anonymous radio-noise sources, should interference be detected.

g. **WIRE SNIFFERS**

Attackers can always insert wire sniffing hardware at cable junctions. It should always be ensured that cable terminals and switch boards are always locked & access be granted only to authorized personnel.

h. **COMPROMISED SERVERS**

An exploited server is a server that is not entirely under your power. Someone else will have gained control of your server, using it for their own motives. Use of a Weak password is often one way a hacker will gain access to your server by guessing your password. People tend to use simple passwords to keep them memorable. Such include dates, lover/pet names, office surrounding etc. Caution must therefore be exercised by combining letters with numerals to create a simple yet strong password.

### i. SERVER SECURITY HOLES

Server Security can be compromised via security holes in a web application like addons/plugins such as joomla / wordpress. It is advisable to use only secure connections whenever possible. This includes the use of SSL connections for email, and SFTP (Secure File transfer Protocol) instead of the more common but unsecure FTP protocol.

### j. ZERO DAY/HOUR ATTACK

Take this for example. The 'sticky keys' feature (sethc.exe) on your XP or Windows7 OS. It is a good accessibility feature that allows one to press special keys only once at a time. This application runs on the logon window when you press shift key five times even before you've entered your password. One only needs to rename the command prompt shell (cmd.exe) to sethc.exe on a logged-in computer. By this, they will have gained full control of your laptop or workspace computer anytime later without passing through any known account. How? By simply pressing the shit key five times and *voila*, the command prompt! Try this for yourself (Hope they got that patched on Windows 8).

Zero hour/day attacks take advantage software vulnerabilities that are yet to catch the eye of a software manufacturer. Should you discover such a bug, report to the software company for a patch to make up for the bug in later releases. Otherwise a hacker may discover the same loophole later, and use it maliciously.

## 3. PROTECTION MODELS

The research community investigated the cyber attack prevention models heavily. Most of the work was focused on preventing such attacks by automating Firewall rules and also improving Access Control Lists on network infrastructure devices. *Alshaer et al.* [3] identified all anomalies that could exist in a single- or multi-firewall environment. They also presented a set of algorithms to detect rule anomalies within a single firewall (intra-firewall anomalies), and between inter-connected firewalls (inter-firewall anomalies) in the network. The authors also presented the Firewall Policy Advisor [22] which provides a number of techniques for purifying and protecting the firewall policy from rule anomalies. The administrator may use the firewall policy advisor to manage firewall policies without prior analysis of filtering rules. In this paper, they formally defined a number of firewall policy anomalies in both centralized and distributed firewalls and they proved that these are the only conflicts that could exist in firewall policies. Then they presented a set of algorithms to detect rule anomalies within a single firewall (intra-firewall anomalies), and between inter-connected firewalls (inter-firewall anomalies) in the network.

The authors in [4] analyzed the local consistency problem in firewall rule sets, with special focus on automatic frequent rule set updates. They also proposed a real time approach to detect inconsistencies in firewall rule sets when inserting, removing or modifying its rules. In [11], the authors approached the problem from a different angle and presented a highly scalable data structure that requires only $O(n)$ space to map the dependencies among firewall policies. Then, they designed an algorithm to iterate over the data structure and detect and eliminate policy conflicts. This proved that the algorithm has an upper bound of $O(n^2 \log n)$, making it the fastest to-date known algorithm for firewall rule anomaly discovery and resolution. They also ran experiments on real-life as well as synthetic firewall policies and showed that their algorithm achieved up to 87% improvement in the number of comparisons overhead, comparatively with the original policies.

*FAME*, Firewall Anomaly Management Environment, [5] is an innovative policy anomaly management framework that facilitates systematic detection and resolution of firewall policy anomalies. It also has a visualization-based firewall policy analysis tool that can used to design policies. In [6], the researchers designed and implemented a firewall analysis tool that allows the administrator to easily discover and test the global firewall policy (either a deployed policy or a planned one). Their tool uses a minimal description of the network topology, and directly parses the various vendor-specific low level configuration files. It interacts with the user through a query-and-answer session.

Alex Liu and his team [7] proposed a framework that can significantly reduce the number of rules in an access control list while maintaining the same semantics, and give an optimal algorithm for the one-dimensional range ACL compression problem, present a systematic solution for compressing multidimensional ACLs with mixed field constraints and conducted extensive experiments on both real-life and synthetic ACLs. Liu and his team also proposed a systematic approach, the TCAM Razor [21], that is effective, efficient, and practical. Systematic approach to minimizing TCAM rules for packet classifiers. While TCAM Razor does not always produce optimal packet classifiers, in their experiments with 40 structurally distinct real-life packet classifier groups, TCAM Razor achieves an average compression ratio of 31.3% and 29.0%, respectively. Unlike other solutions that require modifying TCAM circuits or packet processing hardware, TCAM Razor can be deployed today by network administrators and ISPs to cope with range expansion.

*M. Gouda et al.* [8] proposed a model of stateful firewalls, which is used to store some packets that the firewall has accepted previously and needs to remember in the near future. They designed a model of stateful firewalls that has several favorable properties. It allowed inheriting the rich results in stateless firewall design and analysis. Moreover, it provides backward compatibility such that a stateless firewall can also be specified using our model. Second, they presented methods for analyzing stateful firewalls that are specified using their model.

*Lujo Bauer et al.* [9] showed how to eliminate a large percentage of misconfigurations in advance of attempted accesses using a data-mining technique called association rule mining. Their methods can reduce the number of accesses that would have incurred a costly time-of-access delay by 43%, and can correctly predict 58% of the intended policy.

*B. Hari et al*. [10] proposed a new scheme for conflict resolution, which is based on the idea of adding resolve filters. Their main results are algorithms for detecting and resolving conflicts in a filter database. They have tried their algorithm on 3 existing firewall databases, and have found conflicts, which are potential security holes, in each of them. A general solution is presented for the k -tuple filter, and an optimized version is described for the more common 2-tuple filters consisting of source and destination addresses. They also showed how to use the 2-tuple algorithm for the 5-tuple case in which the other three tuples have a restricted set of values.

*M. Waldvoge et al*. [12] described an algorithm that contains both intellectual and practical contributions. On the intellectual side, after the basic notion of binary searching on hash tables, they found that they had to add markers and use precomputation, to ensure logarithmic time in the worst-case. Algorithms that only use binary search of hash tables are unlikely to provide logarithmic time in the worst case. They single out mutating binary trees as an aesthetically pleasing idea that leverages off the extra structure inherent in their particular form of binary search. On the practical side, they have a fast, scalable solution for IP lookups that can be implemented in either software or hardware. their software projections for IPv4 are 80 ns and they expect 150– 200 ns for IPv6. Our average case speed projections are based on the structure of existing routing databases that they examined. The overall performance can easily be restricted to that of the basic algorithm which already performs

The goal of the work in [13] was to design and implement a high performance, modular, extended integrated services router software architecture in the NetBSD operating system kernel. This architecture allows code modules, called plugins, to be dynamically added and configured at run time. *M. Al-Fares et al*. [14] showed on their paper how to leverage largely commodity Ethernet switches to support the full aggregate bandwidth of clusters consisting of tens of thousands of elements. Similar to how clusters of commodity computers have largely replaced more specialized SMPs and MPPs, they argued that appropriately architected and interconnected commodity switches may deliver more performance at less cost than available from today's higher-end solutions. Their approach requires no modifications to the end host network interface, operating system, or applications; critically, it is fully backward compatible with Ethernet, IP, and TCP.

*M. Abedin et al*. [15] presented an automated process for detecting and resolving such anomalies. The anomaly resolution algorithm and the merging algorithm should produce a compact yet anomaly free rule set that would be easier to understand and maintain. This algorithms can also be integrated into policy advisor and editing tools. They also established the complete definition and analysis of the relations between rules.

*H. Hu et al*. [16] represented an innovative mechanism that facilitates systematic detection and resolution of XACML policy anomalies. A policy-based segmentation technique was introduced to achieve the goals of effective anomaly analysis. Also, described an implementation of a policy anomaly analysis tool called XAnalyzer. The results showed that a policy designer could easily discover and resolve anomalies in an XACML policy with the help of XAnalyzer.

*D. A. Applegate et al.* [20] considered a geometric model for the problem of minimizing access control lists (ACLs) in network routers. Their goal was to create a colored rectilinear pattern within an initially white rectangular canvas, and the basic operation is to choose a subrectangle and paint it a single color, overwriting all previous colors in the rectangle. Rectangle Rule List (RRL) minimization is the problem of finding the shortest list of rules needed to create a given pattern. They provide several equivalent characterizations of the patterns achievable using strip-rules and present polynomial-time algorithms for optimally constructing such patterns when, as in the ACL application, the only colors are black and white (permit or deny). They also showed that RRL minimization is NP-hard in general and provide $O(\min(n1=3; OPT1=2))$- approximation algorithms for general RRL and ACL minimization by exploiting our results about strip-rule patterns. This work was very substantial but it didn't address, however, the integrity of router's Access Control Lists. Consequently, *Ahmat* and *Elnour* [17] investigated the integrity of routers' ACLs in large enterprise networks. More specifically, they studied the problem of discovering and eliminating redundant ACLs from multiple routers' configurations and described efficient methods for removing such redundancies. They also implemented the algorithms they proposed and validated their practicality showing that their approach can discover potential security holes in complex network infrastructures.

*Y. Bartal et al.* [18] presented an initial design and implementation of a prototype for a new generation of firewall and security management tools that showed its usefulness on a real world example, demonstrating that the task of firewall and security configuration/management can be done successfully at a level of abstraction analogous to modern programming languages, rather than assembly code; as an important first step towards the convergence of security and network management. Later, *M. Gritter et al.* [19] described a content routing design based on name-based routing as part of an explicit Internet content layer. The content routing is a natural extension of current Internet directory and routing systems, allows efficient content location, and can be implemented to scale with the Internet. Their results indicate that client name lookup is then faster and far less variable.

## 4. CONCLUSION

In conclusion, a dedicated network security organ is vital for protecting infrastructures. If you have good network security, your company or organization is protected against interruption, employees remain productive. Network security helps you meet compulsory regulatory compliance. Protecting your client's data means no lawsuits emanating from cases about data theft.

Components of the dedicated security organ include:

- A constantly updated Anti-virus software.
- A Firewall, that blocks unauthorized access to workstation PCs (USB ports, LAN, WIFI).
- Virtual Private Networks (VPNs), to give secure remote admission.

## 5. REFERENCES

[1] **A. Alvare,** "How Crackers Crack Passwords or What Passwords to Avoid", Proceedings, UNIX Security Workshop II, August 1990.

[2] **R. Bace, R., and P. Mell, "**Intrusion Detection Systems", NIST Special Publication SP 800-31, November 2000.

[3] **E. Al-Shaer and H. Hamed**, "Discovery of policy anomalies in distributed firewalls", in Proc. IEEE INFOCOM, Mar. 2004, pp. 2605–2616.

[4]  **S. Pozo, R. Ceballos, R. M. Gasca, A. J. Varela-Vaca** "Fast Algorithms for Local Inconsistency Detection in Firewall ACL Updates", 1st International Workshop on Dependability and Security in Complex and Critical Information Systems (DEPEND). Cap Esterel, France. IEEE Computer Society Press, 2008.

Network Security                                    Technical Report – CSE-101507[5] **H. Hu, G.-J. Ahn, and K. Kulkarni**. "Detecting and resolving firewall policy anomalies", IEEE Transactions on Dependable and Secure Computing, 9:318–331, 2012.

[6] **A. Mayer, A. Wool and E. Ziskind**. "Fang: A Firewall Analysis Engine", Proceedings of 2000 IEEE Symposium on Security and Privacy, May 2000.

[7] **Alex X. Liu, Eric Torng, and Chad R. Meiners**. "Compressing network access control lists", IEEE Trans. Parallel Distrib. Syst., 22(12):1969{1977, December 2011. ISSN 1045-9219. doi: 10.1109/TPDS.2011.114. URL http://dx.doi.org/10.1109/TPDS.2011.114.

[8] **M.G. Gouda, A.X. Liu**, "A model of stateful firewalls and its properties", in: Proceedings of the IEEE International Conference on Dependable Systems and Networks (DSN-05), 2005, pp. 320–327.

[9] **S. Garriss, L. Bauer, and M. K. Reiter**. "Detecting and resolving policy misconfigurations in access-control systems", In Proc. of the 13th ACM Symposium on Access Control Models and Technologies, pages 185–194, Estes Park, CO, June 2008.

[10] **B. Hari, S. Suri and G. Parulkar**. "Detecting and Resolving Packet Filter Conflicts", Proceedings of IEEE INFOCOM'00, March 2000.

[11] **H. Gobjuka and K. Ahmat**, "Fast and Scalable Method for Resolving Anomalies in Firewall Policies", in The 14th IEEE Global Internet Symposium (In conjunction with the 31st IEEE International Conference on Computer Communications (INFOCOM 2011), Shanghai, China, 2011.

[12] **M. Waldvoge et al**., "Scalable High Speed IP Routing Lookups", Proc. ACM SIGCOMM '97, Cannes, France, Sept. 1997; http://www.acm.org/sigcomm/ sigcomm97.

[13] **D. Decasper, Z. Dittia, G. Parulkar, B. Plattner**, "Router Plugins: A Modular and Extensible Software Framework for Modern High Performance Integrated Services Routers", TR WUCS-98-08, ARL, Washington University, February 1998.
11 | 1 2

[14] **M. Al-Fares, A. Loukissas, and A. Vahdat. "**A scalable, commodity data center network architecture". In SIGCOMM, 2008.

[15] **Muhammad Abedin, Syeda Nessa, Latifur Khan, Bhavani Thuraisingham**. "Detection and Resolution of Anomalies in Firewall Policy Rules", In Proc. 20th IFIP WG 11.3 Working Conference on Data and Applications Security (DBSec 2006), Springer-Verlag, July 2006, SAP Labs, Sophia Antipolis, France (2006).

[16] **H. Hu, G. Ahn, and K. Kulkarni**. "Anomaly discovery and resolution in web access control policies", In Proceedings of the 16th ACM symposium on Access control models and technologies, pages 165–174. ACM, 2011.

[17] **K. A. Ahmat and A. Elnour**, "Towards Effective Integrated Access Control Lists in Internet Networks", The International Conference on Security and Management (SAM), Las Vegas, Nevada, 2012.

[18] **Y. Bartal., A. Mayer, K. Nissim and A. Wool**. "Firmato: A Novel Firewall Management Toolkit", Proceedings of 1999 IEEE Symposium on Security and Privacy, May 1999.

[19] **M. Gritter, D. Cheriton**, "An Architecture for Content Routing Support in the Internet", Proc. Usenix USITS, March 2001

[20] **D. A. Applegate, G. Calinescu, D. S. Johnson, H. Karloff, K. Ligett, and J.Wang**, "Compressing rectilinear pictures and minimizing access control lists", in Proc. ACM-SIAM SODA, Jan. 2007, pp. 1066–1075.

[21] **Liu, Alex X., Chad R. Meiners, and Eric Torng**. "TCAM Razor: A Systematic Approach Towards Minimizing Packet Classifiers in TCAMs", IEEE/ACM TRANSACTIONS ON NETWORKING 18.2 (2010).

[22] **E. Al-Shaer, H. Hamed, R. Boutaba, and M. Hasan**, "Conflict Classification and Analysis of Distributed Firewall Policies", IEEE J. Sel. Areas Commun., vol. 23, no. 10, pp. 2069–2084, 2005.